\documentclass[preprint,showpacs,preprintnumbers,amsmath,amssymb]{revtex4}
\usepackage{graphicx}

\begin{document}
\thispagestyle{empty}
\title{Constraints on non-Newtonian gravity from measuring
the Casimir force in a configuration with nanoscale
rectangular corrugations}

\author{V.~B.~Bezerra,${}^{1}$
G.~L.~Klimchitskaya,${}^{1,2}$
V.~M.~Mostepanenko,${}^{1,3}$
and C.~Romero${}^{1}$
}

\affiliation{
${}^{1}$Department of Physics, Federal University of Para\'{\i}ba,
C.P.5008, CEP 58059--900, Jo\~{a}o Pessoa, Pb-Brazil \\
${}^{2}${North-West Technical University,
Millionnaya Street 5, St.Petersburg,
191065, Russia}\\
${}^3${Noncommercial Partnership ``Scientific Instruments'',
Tverskaya Street 11, Moscow,
103905, Russia}
}
\begin{abstract}
We report constraints on the parameters of Yukawa-type
corrections to Newtonian gravity from measurements of the
gradient of the Casimir force in the configuration of an
Au-coated sphere above a Si plate covered with corrugations
of trapezoidal shape. For this purpose, the exact
expression for the gradient of Yukawa force in the
experimental configuration is derived and compared with
that obtained using the proximity force approximation.
The reported constraints are of almost the same strength as
those found previously from several different experiments
on the Casimir force and extend over a wide interaction
range from 30 to 1260\,nm. It is discussed how to make them
stronger by replacing the material of the plate.
\end{abstract}
\maketitle

\section{Introduction}

Many extensions of the standard model predict corrections of
Yukawa-type to Newton's gravitational law \cite{1}.
These corrections can be caused by the exchange of light
elementary particles between pairs of atoms belonging to
two macrobodies \cite{2}. Multi-dimensional compactification
schemes, where extra dimensions are compactified at relatively
low energy of the order of 1\,TeV \cite{3,4,5}, also generate
Yukawa-type corrections to Newton's law \cite{6,7}.
Constraints on the parameters of Yukawa-type potential,
the interaction strength $\alpha$ and the interaction range
$\lambda$, were traditionally obtained from the E\"{o}tvos-
and Cavendish-type experiments (see Refs.~\cite{1,8,9} for
a review). For $\lambda$ above a few micrometers the
gravitational experiments resulted in strongest constraints on
$\alpha$ which were used for constraining the masses of predicted
light elementary particles and other parameters of the theory
of fundamental interactions.

For $\lambda$ below a few micrometers, Newtonian gravity becomes
much smaller than the Casimir force induced by electromagnetic
fluctuations. Because of this, the strongest constraints on
non-Newtonian gravity of Yukawa-type in this interaction range
follow from experiments on measuring the Casimir force
(see Ref.~\cite{10} for a review of recent measurements).
As a result, previously known constraints on $\alpha,\,\lambda$
in the submicrometer range were strengthened up to 10000 times
(review of all results obtained from measurements of the
Casimir force before 2009 can be found in Ref.~\cite{11}).
Recently it was also shown that the measure of agreement
between precise measurements of the lateral Casimir force in the
configuration of sinusoidally corrugated surfaces and exact
theory based on the scattering approach \cite{12,13}
leads
to further strengthening of constraints on $\alpha$ in the
interaction range $1.6\,\mbox{nm}\leq\lambda\leq 14\,$nm
\cite{13a}
(note that for $\lambda$ below 1\,nm the strongest constraints on
the Yukawa interactions are obtained \cite{14} from precision
atomic physics).
There was also recent proposal \cite{15} to measure the gradient
of the Casimir force between a plate and a microfabricated
cylinder attached to a micromachined oscillator. It was shown
\cite{16} that this experiment is expected to obtain up to 70
times stronger constraints on $\alpha$ over a wide interaction
range from 12.5 to 630\,nm.

In the present paper we obtain constraints on the parameters of
Yukawa-type corrections to Newtonian gravity resulting from recent
measurements \cite{17} of the gradient
of the Casimir force between a smooth Au-coated sphere and a
Si plate covered with nanoscale rectangular corrugations.
This is the first measurement of the Casimir force in the
configuration with rectangular corrugations where the experimental
data were compared with the exact theory and the measure
of agreement was quantified over some range of separations between
the test bodies. We calculate the gradient of the Yukawa force
acting between an Au-coated sphere and a plate with rectangular
corrugations and derive respective constraints determined by
the magnitude of this measure. Although the derived constraints
are a bit weaker than those of Refs.~\cite{18,19,20},
they serve as a confirmation of previous results obtained
using quite different experimental configuration and
theoretical approach. We also discuss the modifications in the
experimental setup used allowing to significantly strengthen
constraints on the Yukawa interaction obtainable in the
configuration with rectangular corrugated boundary surface.

In Sec.~II we briefly describe the parameters of the
experimental setup \cite{17} needed for obtaining constraints
and derive the measure of agreement between experiment and theory.
In Sec.~III the derivation of the gradient of Yukawa force in
the experimental configuration is presented. This is used
in Sec.~IV to
obtain constraints which are compared with the results following
from other experiments. Prospective constraints obtainable
in the configuration with a rectangular corrugated plate are
considered. Our conclusions and discussion are contained
in Sec.~V.

\section{Measuring the gradient of the Casimir force between
an A$\mbox{u}$ sphere and a S$\mbox{i}$ plate with rectangular corrugations}

The experimental setup of Ref.~\cite{17} includes a micromechanical
torsional oscillator with two attached Au-coated spheres of
$R=50\,\mu$m radii. The thickness of the Au layer on the spheres was
$\Delta_{\rm Au}=400\,$nm.
This oscillator was used in the dynamic regime to measure the
shift of its resonant frequency which is proportional to the
gradient of the Casimir force acting between the upper sphere and
the Si plate covered with rectangular corrugations. In so doing
the distance between both surfaces was measured with high
precision. The corrugations were not precisely rectangular, but
rather had a trapezoidal form shown in Fig.~1 with the following
values of parameters \cite{17}:
$l_1=185.3\pm 2.4\,$nm, $l_2=199.1\pm 4.2\,$nm,
$H=97.8\pm 0.7\,$nm, and $\Lambda=400\,$nm.
The separation $a$ is defined between the mean level of corrugations
and the bottom point of the sphere closest to the Si surface
(in Ref.~\cite{17} separations are defined as $z\approx a-50\,$nm
and the corrugated surface is actually situated above the sphere).
This experiment is the second in a series of measurements using
a rectangular corrugated plate, but the first \cite{22} was not
quantitatively compared with theory.

In Fig.~3(d) of Ref.~\cite{17} the measurement results with the
respective absolute errors as a function of separation were
presented for the quantity
\begin{equation}
\rho(a)=\frac{F_{\rm expt}^{\prime}(a)}{F_{\rm PFA}^{\prime}(a)}.
\label{eq1}
\end{equation}
\noindent
Here $F_{\rm expt}^{\prime}(a)$ is the measured gradient of the
Casimir force between the sphere and the corrugated plate, and
$F_{\rm PFA}^{\prime}(a)\equiv\partial F_{\rm PFA}/\partial a$
is the same gradient calculated using the proximity force
approximation (PFA) \cite{22a,23}. According to the general
formulation of the PFA \cite{22a}, the Casimir force between
a sphere and a periodically corrugated plate can be
approximately represented by
\begin{equation}
F_{\rm PFA}=\frac{1}{\Lambda}\int_{0}^{\Lambda}F\left(z(x)\right)\,dx,
\label{eq2}
\end{equation}
\noindent
where $F\left(z(x)\right)$ is the Casimir force acting between the sphere
and a flat element of the corrugated plate spaced at a separation
$z(x)$ from the sphere.
Using Fig.~1, Eq.~(\ref{eq2}) can be identically rearranged in
the form
\begin{equation}
F_{\rm PFA}(a)=p_1F(a-H_1)+p_2F(a+H_2)
+\frac{2}{\Lambda}
\int_{0}^{l_3}F\left(a+H_2-x\frac{H}{l_3}\right)\,dx,
\label{eq3}
\end{equation}
\noindent
where $p_i=l_i/\Lambda$ ($i=1,\,2$), $l_3=(\Lambda-l_1-l_2)/2$,
and $H_i$ are the distances between the top and the bottom
surfaces of corrugations and the mean level shown by the
dashed line in Fig.~1:
\begin{equation}
H_i=\frac{\Lambda-l_i}{2\Lambda-l_1-l_2}.
\label{eq4}
\end{equation}
\noindent
Equation (\ref{eq3}) is equivalent to Eq.~(1) of Ref.~\cite{17}.
For our purposes, however, it can be further simplified.

In application to a sphere above a flat plate at a separation much smaller than
the sphere radius, the simplified formulation of the PFA \cite{23}
results in
\begin{equation}
F(z)=2\pi RE(z),
\label{eq5}
\end{equation}
\noindent
where $E(z)$ is the Casimir energy per unit area of two plane parallel
material plates expressed by means of the Lifshitz formula \cite{24}.
Calculating the derivative of Eq.~(\ref{eq3}) with respect to the
separation, and taking into account Eq.~(\ref{eq5}) leads to
\begin{equation}
F_{\rm PFA}^{\prime}(a)=2\pi R\left[
\vphantom{\frac{2}{\Lambda}\int_{0}^{l_3}}
-p_1P(a-H_1)-p_2P(a+H_2)
+\frac{2}{\Lambda}
\int_{0}^{l_3}\frac{\partial}{\partial a}
E\left(a+H_2-x\frac{H}{l_3}\right)\,dx\right],
\label{eq6}
\end{equation}
\noindent
where $P(z)$ is the Casimir pressure between the two parallel
material plates expressed by the respective Lifshitz
formula \cite{24}. The integral on the right-hand side of
Eq.~(\ref{eq6}) can be calculated taking into account that
\begin{equation}
\frac{\partial}{\partial a}
E\left(a+H_2-x\frac{H}{l_3}\right)=
-\frac{l_3}{H}\frac{\partial}{\partial x}
E\left(a+H_2-x\frac{H}{l_3}\right).
\label{eq7}
\end{equation}
\noindent
The result for the gradient of the Casimir force is
\begin{equation}
F_{\rm PFA}^{\prime}(a)=-2\pi R\left\{
\vphantom{\frac{1-p_1-p_2}{H}}
p_1P(a-H_1)+p_2P(a+H_2)
+\frac{p_3}{H}\left[
E(a-H_1)-E(a+H_2)\right]\right\},
\label{eq8}
\end{equation}
\noindent
where $p_3=1-p_1-p_2$.

Measurements of the gradient of the Casimir force were performed
over the separation region from 200 to 550\,nm.
As is shown in Sec.~IV below, the strongest constraints on the
parameters of Yukawa-type interaction follow from the measurement
data at separations 279, 384, and 413\,nm.
{}From Fig.~3(d) of Ref.~\cite{17} it can be seen that the
absolute errors $\Delta\rho$ of the quantity $\rho$ at these
separations are equal to 0.025, 0.022, and 0.029, respectively.
Using the definition (\ref{eq1}) and Eq.~(\ref{eq8}) for
$F_{\rm PFA}^{\prime}(a)$, the errors $\Delta\rho$ can be
recalculated to give the absolute errors
$\Delta F_{\rm expt}^{\prime}(a)$ of the gradient of the
Casimir force. In doing so the energy per unit area of parallel
plates, $E$, and the Casimir pressure, $P$, were calculated
using the Lifshitz formula taking into account the surface
roughness by means of geometrical averaging \cite{10}
(rms roughness on the sphere was measured to be
of about 4\,nm and on the plate of about 0.6\,nm \cite{17}).
Following Ref.~\cite{17}, the dielectric properties of Au
were described using the tabulated optical data \cite{25a}
extrapolated to low frequencies by means of the Drude model
with the plasma frequency $\omega_p=9\,$eV and relaxation
parameter $\gamma=35\,$meV. For Si, the Drude-Lorentz form of
dielectric permittivity was used. As shown in Refs.~\cite{27a,27},
computations using this theoretical approach are in a very
good agreement with the experimental data. The error in the
gradient of the Casimir force arising from the errors in the
optical data was estimated to be of about 0.5\% \cite{10}.
As a result, we have determined that the absolute errors of
the measured gradient of the Casimir force
$\Delta F_{\rm expt}^{\prime}(a)$ were equal to
$9.0\times 10^{-7}$,  $2.0\times 10^{-7}$, and
$1.9\times 10^{-7}\,$N/m at separations 279, 384, and 413\,nm,
respectively.

Recent progress in the theory of the Casimir force,
pioneered by Ref.~\cite{25}, allowed exact computations
between surfaces of arbitrary shape. In Ref.~\cite{17}
the experimental data were found to be in agreement with
exact computational results within the limits of
experimental errors at separations from 279 to 530\,nm.
Exact computations were performed at zero temperature
using the same dielectric functions for Au and Si, as
indicated above. Errors in knowledge of surface profile
do not play an important role in the accuracy of
computed gradient of the Casimir force. They are more
influencial for the computation of Yukawa force (see
Sec.~III).
At shorter separations the experimental data (five data
points) were found to disagree with the exact theory
within the errors indicated in Fig.~3(d) of Ref.~\cite{17}.
This might be explained by the slow convergence of
computations in the exact theory and insufficient number
of iterations used. Below we obtain constraints on the
parameters of Yukawa-type interaction using only the
separation region where the measurement data are in
good agreement with the exact theory.

\section{Yukawa-type force
between a sphere and a plate with rectangular corrugations}

The Yukawa interaction potential between two point masses
$m_1$ and $m_2$ (atoms or molecules) at a separation $r$
is usually represented by \cite{1,8,9,11}
\begin{equation}
V_{\rm Yu}(r)=-\frac{Gm_1m_2\alpha}{r}\,
e^{-r/\lambda},
\label{eq9}
\end{equation}
\noindent
where $G$ is the Newtonian gravitational constant.
We assume that $m_1$ belongs to an Au-coated sphere and
$m_2$ to a Si plate covered with rectangular corrugations
shown in Fig.~1. The Yukawa force between these test bodies
can be obtained by the integration of Eq.~(\ref{eq9}) over
their volumes and subsequent negative differentiation with
respect to the separation $a$. Most simply it can be obtained
in the following way.

First we consider the previously derived \cite{26} Yukawa
force between the sphere of density $\rho_s$ covered with
an Au layer with density $\rho_{\rm Au}$ of thickness
$\Delta_{\rm Au}$ and a large thick plate (semispace) of
density $\rho_p$ at a separation $a$ from the bottom point of
the sphere. The modelling of a plate by the semispace is well
justified because the side length of a plate is of about
$700\,\mu$m \cite{22}, i.e., much larger than the sphere
radius. As to the plate thickness, it can be considered as
infinitely large even for plates of $10\,\mu$m thickness
when the Yukawa-type interactions with $\lambda<1\,\mu$m
are considered (in reality, experiments on measuring the
Casimir force employ Si plates of order $100\,\mu$m
thickness \cite{27}).
In this case the exact expression for the Yukawa force is given
by \cite{26}
\begin{equation}
F_{{\rm Yu},sp}(a)=-4\pi^2 R\alpha G\lambda^3e^{-a/\lambda}
\rho_p\left[
\vphantom{e^{-\Delta_{\rm Au}/\lambda}}
\rho_{\rm Au}\Phi_1(\lambda/R)
-(\rho_{\rm Au}-\rho_s)\Phi_2(\lambda/R,\Delta_{\rm Au}/R)
e^{-\Delta_{\rm Au}/\lambda}\right],
\label{eq10}
\end{equation}
\noindent
where the functions $\Phi_1(x)$ and $\Phi_2(x,y)$ are
defined as follows:
\begin{eqnarray}
&&
\Phi_1(x)=1-x+(1+x)e^{-2/x},
\label{eq11} \\
&&
\Phi_2(x,y)=1-x-y+(1+x-y)e^{-2(1-y)/x}.
\nonumber
\end{eqnarray}
\noindent
Note that in the asymptotic limit
\begin{equation}
x=\frac{\lambda}{R}\ll 1
\quad\mbox{and}\quad
y=\frac{\Delta_{\rm Au}}{R}\ll 1
\label{eq11a}
\end{equation}
\noindent
we have $\Phi_{1,2}\to 1$ and Eq.~(\ref{eq10}) takes the
form
\begin{equation}
F_{{\rm Yu},sp;{\rm PFA}}(a)=2\pi RE_{\rm Yu}(a),
\label{eq12}
\end{equation}
\noindent
where
\begin{equation}
E_{\rm Yu}(a)=-2\pi\alpha G\lambda^3e^{-a/\lambda}
\rho_p\left[\rho_{\rm Au}
-(\rho_{\rm Au}-\rho_s)
e^{-\Delta_{\rm Au}/\lambda}\right].
\label{eq13}
\end{equation}
\noindent
The quantity $E_{\rm Yu}(a)$ in Eq.~(\ref{eq13}) coincides
\cite{28} with the energy of Yukawa interaction per unit area
of two semispaces separated with a gap of width $a$, one of
which is homogeneous and has density $\rho_p$ and the
other one has density $\rho_s$ and is covered by the outer
layer of density $\rho_{\rm Au}$ and thickness $\Delta_{\rm Au}$.
Comparing Eq.~(\ref{eq12}) with Eq.~(\ref{eq5}), one can conclude
that under the conditions (\ref{eq11a}) the Yukawa-type force
between an Au-coated sphere and a semispace can be calculated
by using the simplified formulation of the proximity force
approximation.

As the next step we calculate the exact Yukawa force between
an Au-coated sphere and a corrugated plate shown in Fig.~1.
Similar to the case of the Casimir force,
this can be presented in the form
\begin{equation}
F_{\rm Yu,corr}(a)=\frac{1}{\Lambda}
\int_{0}^{\Lambda}F_{{\rm Yu},sp}\left(z(x)\right)\,dx,
\label{eq14}
\end{equation}
\noindent
where
$F_{{\rm Yu},sp}$ is defined in Eq.~(\ref{eq10}).
Note, however, that Eq.~(\ref{eq2}) for the Casimir force is an
approximate one, whereas Eq.~(\ref{eq14}) for the Yukawa force
is the exact expression.
Taking into account the explicit expression for $z(x)$,
Eq.~(\ref{eq14}) can be  rearranged to give
\begin{equation}
F_{\rm Yu,corr}(a)=p_1F_{{\rm Yu},sp}(a-H_1)+p_2F_{{\rm Yu},sp}(a+H_2)
+\frac{2}{\Lambda}
\int_{0}^{l_3}F_{{\rm Yu},sp}\left(a+H_2-x\frac{H}{l_3}\right)\,dx.
\label{eq15}
\end{equation}
\noindent
The integral on the right-hand side of Eq.~(\ref{eq15}) is
calculated taking into account Eq.~(\ref{eq10})
\begin{equation}
\frac{2}{\Lambda}
\int_{0}^{l_3}F_{{\rm Yu},sp}\left(a+H_2-x\frac{H}{l_3}\right)\,dx
=\frac{p_3\lambda}{H}
\left[F_{{\rm Yu},sp}(a-H_1)
-F_{{\rm Yu},sp}(a+H_2)\right].
\label{eq16}
\end{equation}
\noindent
{}From Eqs.~(\ref{eq15}) and (\ref{eq16}), for the gradient of the
Yukawa force between an Au-coated sphere and a corrugated plate
we obtain
\begin{equation}
F_{\rm Yu,corr}^{\prime}(a)=\left(p_1+p_3\frac{\lambda}{H}\right)
F_{{\rm Yu},sp}^{\prime}(a-H_1)
+\left(p_2-p_3\frac{\lambda}{H}\right)
F_{{\rm Yu},sp}^{\prime}(a+H_2).
\label{eq17}
\end{equation}
\noindent
Note that errors in the knowledge of corrugation profiles lead to
respective errors in computed gradient of the Yukawa force.
The largest error arises from the error in $l_1$ which is of
about 1.3\%. This effectively leads to the same error in the
interaction strength $\alpha$.
In the logarithmic scale used in Fig.~2 in Sec.~IV, however, the
resulting maximum relative error of $\log_{10}\alpha$ does not
exceed 0.12\% and, thus, is negligibly small.

In the application region of the PFA (\ref{eq11a}), we can differentiate
Eq.~(\ref{eq12}) with respect to $a$ and obtain the following result:
\begin{equation}
F_{\rm Yu,corr;{\rm PFA}}^{\prime}(a)=2\pi R
E_{\rm Yu}^{\prime}(a)=-2\pi R P_{\rm Yu}(a),
\label{eq18}
\end{equation}
\noindent
where $P_{\rm Yu}(a)$ is the Yukawa pressure between the two plates,
one with density $\rho_p$ and the
other one with density $\rho_s$ is covered with an Au
layer of density $\rho_{\rm Au}$ and thickness $\Delta_{\rm Au}$.
In view of Eq.~(\ref{eq18}), Eq.~(\ref{eq17}) can be rewritten as
\begin{equation}
F_{\rm Yu,corr;{\rm PFA}}^{\prime}(a)=-2\pi R
\left[\left(p_1+p_3\frac{\lambda}{H}\right)
P_{{\rm Yu}}(a-H_1)
+\left(p_2-p_3\frac{\lambda}{H}\right)
P_{{\rm Yu}}(a+H_2)\right].
\label{eq19}
\end{equation}
\noindent

For the Yukawa interaction, the differentiation of Eq.~(\ref{eq13})
leads to the equality
\begin{equation}
\lambda P_{\rm Yu}(z)=E_{\rm Yu}(z).
\label{eq20}
\end{equation}
\noindent
When we substitute Eq.~(\ref{eq20}) into Eq.~(\ref{eq19}) and
compare the obtained result with Eq.~(\ref{eq8}), it becomes
evident that in the application region of the PFA the gradient of
the Yukawa force between the sphere and the corrugated plate is
given by the same mathematical expression as the gradient of
the Casimir force.

\section{Constraints on non-Newtonian gravity from the experiment
with corrugated plate}

The constraints on the parameters of Yukawa-type interactions
$\alpha,\,\lambda$, following from measurements of the Casimir
force between an Au-coated sphere and a Si plate covered with
nanoscale corrugations, can be obtained from the inequality
\begin{equation}
|F_{\rm Yu,corr}^{\prime}(a)|\leq\Delta F_{\rm expt}^{\prime}(a).
\label{eq21}
\end{equation}
\noindent
Here, the absolute errors in the measured gradient of the Casimir
force are indicated in Sec.~II. Taking  Eq.~(\ref{eq10}) into
account, the gradient
of the Yukawa force (\ref{eq17})
can be written in the form
\begin{equation}
F_{\rm Yu,corr}^{\prime}(a)=
-\left(\frac{p_1}{\lambda}+\frac{p_3}{H}\right)
F_{{\rm Yu},sp}(a-H_1)-
\left(\frac{p_2}{\lambda}-\frac{p_3}{H}\right)
F_{{\rm Yu},sp}(a+H_2).
\label{eq22a}
\end{equation}
\noindent
It is easily seen that
the magnitude of this gradient
is given by
\begin{equation}
|F_{\rm Yu,corr}^{\prime}(a)|=
\left(\frac{p_1}{\lambda}+\frac{p_3}{H}\right)
|F_{{\rm Yu},sp}(a-H_1)|
+\left(\frac{p_2}{\lambda}-\frac{p_3}{H}\right)
|F_{{\rm Yu},sp}(a+H_2)|,
\label{eq22}
\end{equation}
\noindent
where $F_{{\rm Yu},sp}(z)$ is defined in Eq.~(\ref{eq10}).

We have obtained the constraints on $\alpha,\,\lambda$ following
from Eqs.~(\ref{eq21}) and (\ref{eq22}) at all separations from
279 to 980\,nm where the measurement results were in agreement
with the exact theory within the limits of experimental errors.
In doing so the following densities of the interacting bodies
were used: $\rho_{\rm Au}=19.28\times 10^3\,\mbox{kg/m}^3$,
$\rho_{p}=2.33\times 10^3\,\mbox{kg/m}^3$, and
$\rho_{s}=2.23\times 10^3\,\mbox{kg/m}^3$.
The strongest constraints were obtained at $a=279\,$nm
(in the interaction range from $\lambda=30\,$nm to
$\lambda=63\,$nm), at $a=384\,$nm
(for $63\,\mbox{nm}<\lambda\leq 500\,$nm), and at $a=413\,$nm
(for $500\,\mbox{nm}<\lambda\leq 1.26\,\mu$m).
These constraints are shown as line 1 in Fig.~2, where
the regions of $(\alpha,\lambda)$-plane above each line are
prohibited and below each line are allowed by the results of
the respective experiments. For comparison purposes, in the same
figure we show constraints on $(\alpha,\lambda)$ obtained
from the measurement of the equivalent Casimir pressure
between two parallel plates by means of micromechanical
torsional oscillator with a flat
plate \cite{19,20} (line 2) and from the
Casimir-less experiment \cite{18} (line 3).
Line 4 shows constraints obtained \cite{28a} from the
experiment \cite{29} with torsion pendulum.
The constraints obtained
from a more recent experiment with torsion pendulum
\cite{21} are shown by line 5.
Finally, line 6 shows the constraints obtained in Ref.~\cite{13a}
from the measurement of the thermal Casimir-Polder
force \cite{29a}.
It is pertinent to note that both experiments \cite{29,21}
performed with a torsion pendulum exploited spherical
lenses of more than 10\,cm radii of curvature and used
the simplified formulation of the PFA (\ref{eq5}) to
compare the measured Casimir force with theory.
Recently it was shown, however, that for lenses of
centimeter-size radii of curvature, Eq.~(\ref{eq5}) is
not applicable for the calculation of the Casimir force
due to surface imperfections, such as bubbles and pits,
which are invariably present on the surfaces of large
lenses \cite{30,31}. Because of this, the constraints on
$(\alpha,\lambda)$ shown by the lines 4 and 5 invite further
investigation and check.

As can be seen in Fig.~2, the constraints shown by line 1
are a bit weaker than the constraints shown by lines 2 and 3.
At $\lambda=1.26\,\mu$m, where the minimum difference between the
constraints shown by lines 2 and 3 from the one hand and
line 1 from the other hand is achieved, our constraint
(line 1) is weaker by a factor of 2.4. This is explained by
the smaller
density of Si in comparison with the density of Au deposited on
both test bodies in the experiments of Refs.~\cite{18,19,20}.
At the same time, line 1 in Fig.~2 is in very good
qualitative agreement with lines 2 and 3 and provides additional
confirmation to previously obtained constraints shown by
lines 2 and 3.

The above constraints were obtained from Eq.~(\ref{eq21}) using
the exact formulas (\ref{eq22}), (\ref{eq10}) for the gradient
of the Yukawa force. It is interesting to compare them with
constraints obtained using a more simple Eq.~(\ref{eq19}) for
the gradient of the Yukawa force using the PFA.
To give a few examples,
the maximum allowed values of $\alpha$ computed at several
interacting ranges using the exact gradient of the Yukawa force
and the gradient of the Yukawa force calculated within the PFA
are the following:
\begin{center}
\begin{tabular}{lll}
$\lambda=10\,\mbox{nm}$, & $\alpha_{\rm ex}=2.8565\times 10^{25}$, &
$\alpha_{\rm PFA}=2.8559\times 10^{25}$,
\\
$\lambda=100\,\mbox{nm}$, & $\alpha_{\rm ex}=1.470\times 10^{14}$, &
$\alpha_{\rm PFA}=1.467\times 10^{14}$,
 \\
$\lambda=1000\,\mbox{nm}$, & $\alpha_{\rm ex}=1.212\times 10^{11}$,&
$\alpha_{\rm PFA}=1.202\times 10^{11}$.
\end{tabular}
\end{center}
\noindent
It is seen that under condition (\ref{eq11a}) the application
of the PFA to calculate the Yukawa force results in almost
the same strength of constraints as obtained using the exact
Yukawa interaction. This provides one more confirmation to the
conclusion of Refs.~\cite{26,28} that in its application range
the PFA is well applicable for calculation of both the Casimir
and Yukawa-type forces.

We emphasize that the experiment of Ref.~\cite{17} under
consideration here leads to competitive constraints on the
parameters of Yukawa-type interaction over a very wide
interaction range from 30\,nm to $1.26\,\mu$m.
This experiment alone covers the interaction ranges of two
earlier performed experiments of Refs.~\cite{19,20} (the
obtained constraints are shown by line 2) and
of Ref.~\cite{18}
(constraints of line 3).
If the material of the corrugated plate (Si) were replaced
by Au, the
constraints shown by the dashed line in Fig.~2 would
be obtained.
These constraints are stronger by a factor of
$\rho_{\rm Au}/\rho_p\approx 8.3$ than the constraints of
line 1.
Within some interaction ranges the dashed line also
demonstrates stronger constraints than those shown by lines
3, 5, and 6. Thus, at $\lambda=0.94\,\mu$m the constraint
shown by the dashed line is by a factor 3.8 stronger
than the constraints shown by lines 3 and 5.
At $\lambda=1.04\,\mu$m the dashed line gives a
constraint which is stronger by a factor 3.4
than that shown by lines 3 and 6.

The above results demonstrate that the experiment of
Ref.~\cite{17} has considerable opportunity for obtaining
constraints on non-Newtonian gravity. Further strengthening
of constraints by means of this experiment can be achieved
by decreasing the experimental errors and increasing the accuracy
of computations of the Casimir force in the framework of
the exact theory.

\section{Conclusions and discussion}

In the foregoing we have obtained constraints on the parameters
of Yukawa-type corrections to Newtonian gravity which follow
from the measurements of the gradient of the Casimir force
between an Au-coated sphere and a corrugated Si plate \cite{17}.
For this purpose the exact expression for the gradient of the
Yukawa force which acts in the experimental configuration of
a sphere interacting with a plate covered with trapezoidal
corrugations was derived. The expression obtained was compared
with that derived using the PFA. The constraints on
non-Newtonian gravity were obtained from the measure of
agreement between the experimental results and the exact
theory of the Casimir force.

According to our results, the obtained constraints are only
a bit weaker than those derived from the two different
experiments using both Au-coated test bodies \cite{18,19,20}.
They cover an extremely wide interaction range from 30\,nm
to 1260\,nm. The same constraints are obtained if the
gradient of the Yukawa force were computed using the PFA.
This confirms the result of Refs.~\cite{26,28} that within
its application range the PFA is well applicable to the
calculation of the Yukawa-type force.

Some ways of
 how to strengthen the obtained constraints on the
parameters of Yukawa-type interaction using the corrugated surfaces
are proposed. It is shown that the use of Au corrugated plate
with the same experimental parameters as for a Si plate in
Ref.~\cite{17} would lead to constraints several times stronger
than the ones obtained from the experiments of
Refs.~\cite{18,19,20,21,29a}.
Further strengthening is possible at the expense of a decrease of
errors in the force measurement and widening the separation
interval where the experimental data are in agreement with
theoretical results.

\section*{Acknowledgments}

The authors were supported by CNPq (Brazil).
G.~L.~K.\ and V.~M.~M.\ are grateful to the Federal University of
Para\'{\i}ba, where this work was
performed, for kind hospitality.


\begin{figure}[h]
\vspace*{-18.3cm}
\centerline{\hspace*{2cm}
\includegraphics{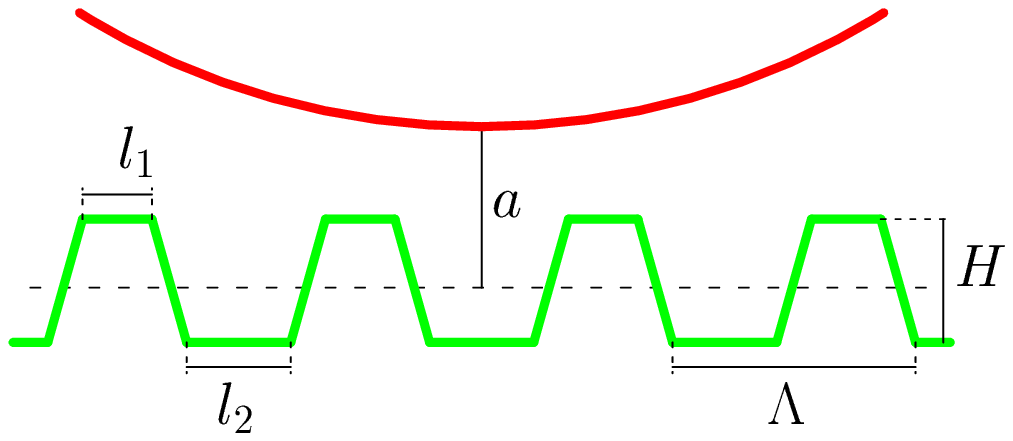}
}
\vspace*{-7.3cm}
\caption{(color online).
The configuration of a sphere and a plate covered with
corrugations of trapezoidal shape
(the relative sizes  are shown not to scale).
See text for the values of all parameters.
}
\end{figure}
\begin{figure}[h]
\vspace*{-16cm}
\centerline{\hspace*{3cm}
\includegraphics{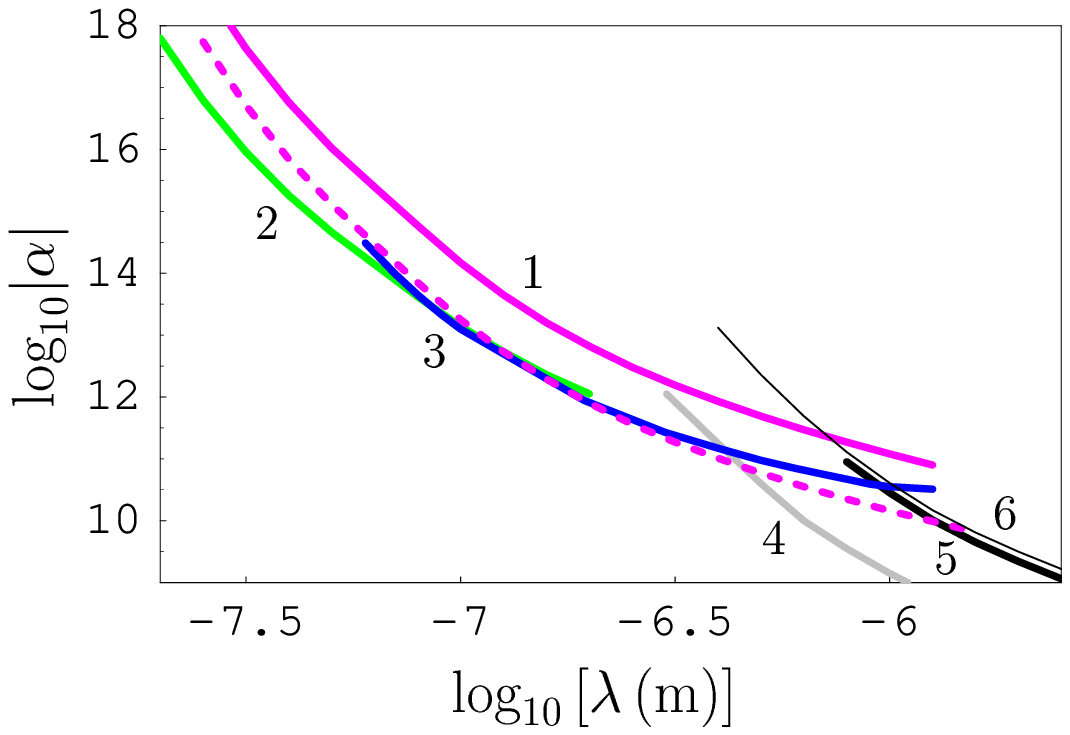}
}
\vspace*{-7cm}
\caption{(color online).
Constraints on the parameters of Yukawa-type interaction
which are obtained from the experiments performed
by means of a micromechanical torsional oscillator
with a corrugated Si plate (line 1) and with a flat
Au-coated plate (line 2), from the Casimir-less experiment
(line 3), from the experiments using a torsion pendulum
(lines 4 and 5) and from measuring the thermal Casimir-Polder
force (line 6). The dashed line shows the prospective constraints
that can be obtained from the experiment with a
corrugated Au plate.
The allowed regions in the $(\lambda,\,\alpha)$-plane lie
beneath the lines.
See text for the references to related experiments.
}
\end{figure}

\begin{thebibliography}{99}
\bibitem{1}
E.~Fischbach and C.~L.~Talmadge, {\it The Search for Non-Newtonian
Gravity} (Springer, New York, 1999).
\bibitem{2}
V.\ M.\ Mostepanenko and I.~Yu.~Sokolov,
Phys. Rev. D {\bf 47}, 2882 (1993).
\bibitem{3}
I.~Antoniadis,
N.~Arkani-Hamed, S.~Dimopoulos, and G.~Dvali,
Phys. Lett. B {\bf 436}, 257 (1998).
\bibitem{4}
N.~Arkani-Hamed, S.~Dimopoulos, and G.~Dvali,
Phys. Lett. B {\bf 429}, 263 (1998).
\bibitem{5}
N.~Arkani-Hamed, S.~Dimopoulos, and G.~Dvali,
Phys. Rev. D {\bf 59}, 086004 (1999).
\bibitem{6}
E.~G.~Floratos and G.~K.~Leontaris,
Phys. Lett. B {\bf 465}, 95 (1999).
\bibitem{7}
A.~Kehagias and K.~Sfetsos,
Phys. Lett. B {\bf 472}, 39 (2000).
\bibitem{8}
E.~G.~Adelberger, B.~R.~Heckel, C.~W.~Stubbs,
and W.~F.~Rogers,
 Ann. Rev. Nucl. Part. Sci. {\bf 41}, 269 (1991).
\bibitem{9}
E.~G.~Adelberger, B.~R.~Heckel, and A.~E.~ Nelson,
 Ann. Rev. Nucl. Part. Sci. {\bf 53}, 77 (2003).
\bibitem {10}
G.~L.~Klimchitskaya, U. Mohideen, and V.\ M.\ Mostepanenko,
Rev. Mod. Phys. {\bf 81}, 1827 (2009).
\bibitem{11}
M.~Bordag, G.~L.~Klimchitskaya, U.\ Mohideen, and
V.\ M.\ Mostepanenko, {\it Advances in the Casimir Effect}
(Oxford University Press, Oxford, 2009).
\bibitem{12}
H.-C.\ Chiu,  G.~L.~Klimchitskaya, V.\ N.\ Marachevsky,
V.\ M.\ Mos\-te\-pa\-nen\-ko, and U.~Mohideen,
Phys. Rev. B {\bf 80}, 121402(R) (2009).
\bibitem{13}
H.-C.\ Chiu,  G.~L.~Klimchitskaya, V.\ N.\ Marachevsky,
V.\ M.\ Mos\-te\-pa\-nen\-ko, and U.~Mohideen,
Phys. Rev. B {\bf 81}, 115417 (2010).
\bibitem{13a}
V.~B.~Bezerra, G.~L.~Klimchitskaya,
 V.~M.~Mostepanenko, and C.~Romero,
Phys. Rev. D {\bf 81}, 055003 (2010).
\bibitem{14}
S.~G.~Karshenboim,
Phys. Rev. D {\bf 81}, 073003 (2010).
\bibitem{15}
R.~S.~Decca,  E.~Fischbach, G.~L.~Klimchitskaya, D.~E.~Krause,
D.~L\'opez, and V.~M.~Mostepanenko,
Phys. Rev. A {\bf 82}, 052515 (2010).
\bibitem{16}
G.~L.~Klimchitskaya and C.~Romero,
Phys. Rev. D {\bf 82}, 115005 (2010).
\bibitem{17}
Y.~Bao, R.~Gu\'{e}rout, J.~Lussange, A.\ Lambrecht, R.\ A.\ Cirelli, F.\ Klemens,
W.\ M.\ Mansfield, C.\ S.\ Pai,
and H.~B.~Chan,
{\it Phys. Rev. Lett.} {\bf 105}, 250402 (2010).
\bibitem{18}
R.~S.~Decca, D.~L\'opez, E.~Fischbach,
 D.~E.~Krause, and C.~R.~Jamell,
Phys. Rev. Lett. {\bf 94}, 240401 (2005).
\bibitem{19}
R.~S.~Decca, D.~L\'opez, E.~Fischbach, G.~L.~Klimchitskaya,
 D.~E.~Krause, and V.~M.~Mostepanenko,
Phys. Rev. D {\bf 75}, 077101 (2007).
\bibitem{20}
R.~S.~Decca, D.~L\'opez, E.~Fischbach, G.~L.~Klimchitskaya,
 D.~E.~Krause, and V.~M.~Mostepanenko,
Eur. Phys. J. C {\bf 51}, 963 (2007).
\bibitem{22}
H.~B.~Chan, Y.~Bao, J.~Zou, R.\ A.\ Cirelli, F.\ Klemens,
W.\ M.\ Mansfield, and C.\ S.\ Pai,
{\it Phys. Rev. Lett.} {\bf 101}, 030401 (2008).
\bibitem{22a}
B.~V.~Derjaguin, Kolloid. Z. {\bf 69}, 155 (1934).
\bibitem{23}
J.~B{\l}ocki, J.~Randrup, W.~J.~Swiatecki, and C.~F.~Tsang,
Ann. Phys. (N.Y.) {\bf 105}, 427 (1977).
\bibitem{24}
E.~M.~Lifshitz and L.~P.~Pitaevskii,
{\it Statistical Physics}, Part.~II (Pergamon Press, Oxford, 1980).
\bibitem{25a}
{\it Handbook of Optical Constants of Solids},
ed. E.~D.~Palik (Academic, New York, 1985).
\bibitem{27a}
F.~Chen, U.~Mohideen, G.~L.~Klimchitskaya,
and V.\ M.\ Mos\-te\-pa\-nen\-ko,
Phys. Rev. A {\bf 74}, 022103 (2006).
\bibitem{27}
F.~Chen, G.~L.~Klimchitskaya,
V.\ M.\ Mos\-te\-pa\-nen\-ko, and  U.~Mohideen,
Phys. Rev. Lett. {\bf 97}, 170402 (2006).
\bibitem{25}
T.~Emig, R.~L.~Jaffe, M.~Kardar, and A.~Scardicchio,
Phys. Rev. Lett. {\bf 96}, 080403  (2006).
\bibitem{26}
E.~Fischbach, G.~L.~Klimchitskaya,
 D.~E.~Krause, and V.~M.~Mostepanenko,
Eur. Phys. J. C {\bf 68}, 223 (2010).

\bibitem{28}
R.~S.~Decca,  E.~Fischbach, G.~L.~Klimchitskaya,
 D.~E.~Krause, D.~L\'opez, and V.~M.~Mostepanenko,
Phys. Rev. D {\bf 79}, 124021 (2009).
\bibitem{28a}
M.~Bordag, B.~Geyer, G.~L.~Klimchitskaya,
and V.~M.~Mostepanenko,
{Phys. Rev. D}  {\bf 58}, 075003 (1998).
\bibitem{29}
S.~K.~Lamoreaux, { Phys. Rev. Lett.} {\bf 78}, 5 (1997);
{\bf 81}, 5475(E) (1998).
\bibitem{21}
M.~Masuda and M.~Sasaki,
Phys. Rev. Lett. {\bf 102}, 171101  (2009).
\bibitem{29a}
J.~M.~Obrecht, R.~J.~Wild, M.~Antezza, L.~P.~Pitaevskii,
S.~Stringari, and E.~A.~Cornell,
Phys. Rev. Lett. {\bf 98}, 063201 (2007).
\bibitem{30}
G.~L.~Klimchitskaya and V.~M.~Mostepanenko,
e-print arXiv:1010.2216v1.
\bibitem{31}
V.~B.~Bezerra,
G.\ L.\ Klimchitskaya, U.~Mohideen,
V.\ M.\ Mostepanenko, and C.\ Romero,
Phys. Rev. B {\bf 83}, 075417 (2011).



\end{thebibliography}
\end{document}